\documentclass{jetpl}
\twocolumn
\usepackage{graphicx}
\usepackage{bm}
\lat
\usepackage[usenames,dvipsnames]{color} 
\newcommand{\mkrm}[1]{}           

\title{Phonon-particle coupling effects in odd-even mass differences of semi-magic nuclei.}%

\rtitle{Phonon-particle coupling effects in odd-even mass differences  of semi-magic nuclei.}%

\sodtitle{Phonon-particle coupling effects in odd-even mass differences  of semi-magic nuclei.}%

\author{
 E.\,E.\,Saperstein$^{*,**}$\/\thanks{e-mail: Sapershtein\_EE@nrcki.ru},
 M.\,Baldo$^{\dagger}$,
 S.\,S.\,Pankratov$^{*,***}$,
 S.\,V.\,Tolokonnikov$^{*,***}$}

\rauthor{E.\,E.\,Saperstein, M.\,Baldo, S.\,S.\,Pankratov, S.\,V.\,Tolokonnikov}

\sodauthor{E.\,E.\,Saperstein, M.\,Baldo, S.\,S.\,Pankratov, S.\,V.\,Tolokonnikov}

\address{$^{*}$National Research Centre Kurchatov Institute, pl.
Akademika Kurchatova 1, Moscow, 123182 Russia\\
$^{**}$National Research Nuclear University MEPhI, 115409 Moscow, Russia\\
$^{\dagger}$INFN, Sezione di Catania, 64 Via S.-Sofia, I-95125 Catania, Italy\\
$^{***}$Moscow Institute of Physics and Technology, 141700 Dolgoprudny, Russia}

\dates{\today}{*}

\abstract{A method to evaluate the particle-phonon coupling (PC) corrections to the single-particle
energies in  semi-magic nuclei, based on a direct solving the Dyson equation with PC corrected mass
operator, is used for finding the odd-even mass difference between 18 even Pb isotopes and their
odd-proton neighbors.  The Fayans energy density functional (EDF) DF3-a is used which gives rather
high accuracy of the predictions for these mass differences already on the mean-field level, with the
average deviation from the existing experimental data equal to 0.389 MeV. It is only a bit worse than
the corresponding value of 0.333 MeV  for the Skyrme EDF HFB-17 which belongs to a family of Skyrme
EDFs with the highest overall accuracy in describing the nuclear masses. Account for the PC
corrections induced by the low-laying phonons $2^+_1$ and $3^-_1$ significantly diminishes the
deviation of the theory from the data till 0.218 MeV.}

\PACS{21.60.-n; 21.65.+f; 26.60.+c; 97.60.Jd}

\begin{document}

\newcommand{\beq}{\begin{equation}}
\newcommand{\eeq}{\end{equation}}
\newcommand{\bea}{\begin{eqnarray}}
\newcommand{\eea}{\end{eqnarray}}
\newcommand{\eps}{\varepsilon}
\newcommand{\bfg}{\boldsymbol}

\maketitle

The single-particle (SP) spectrum of a nucleus essentially influences different nuclear properties.
Therefore ability to describe nuclear SP spectra correctly is very important for any self-consistent
nuclear theory. Till now, experimental SP levels are known in detail only for magic nuclei
\cite{lev-exp}. Their analysis in \cite{levels} on the base of the energy density functional (EDF)
DF3-a \cite{DF3-a}, which is a small modification of the original Fayans EDF DF3 \cite{Fay1,Fay}, led to rather good
description of the data, significantly better than the predictions of the popular Skyrme EDF HFB-17
\cite{HFB-17} which belongs to the family of the EDFs HFB-17 -- HFB-27 \cite{HFB-site} possessing the highest accuracy
among all the self-consistent calculations in reproducing nuclear masses.

Inclusion of the
particle-phonon coupling (PC) corrections, with account for the non-pole diagrams
\cite{Khod-76,scTFFS}, made the agreement even better. So-called $g_L^2$-approximation for the PC
corrections was applied, $g_L$ being the creation vertex of the $L$-phonon. The corresponding diagrams
for the PC correction $\delta \Sigma^{\rm PC}_L$ to the mass operator $\Sigma_0$, in the
representation of the SP states $|\lambda\rangle$, are displayed in Fig. 1. The first one is the usual
pole diagram, with obvious notation, whereas the second one represents the sum of all non-pole
diagrams of the order $g_L^2$. The latter is often named ``the phonon tadpole'' \cite{phon-tad}, as an
analogue of the tadpole-like diagrams in the field theory \cite{tad}.

\begin{figure}
\centerline {\includegraphics [width=80mm]{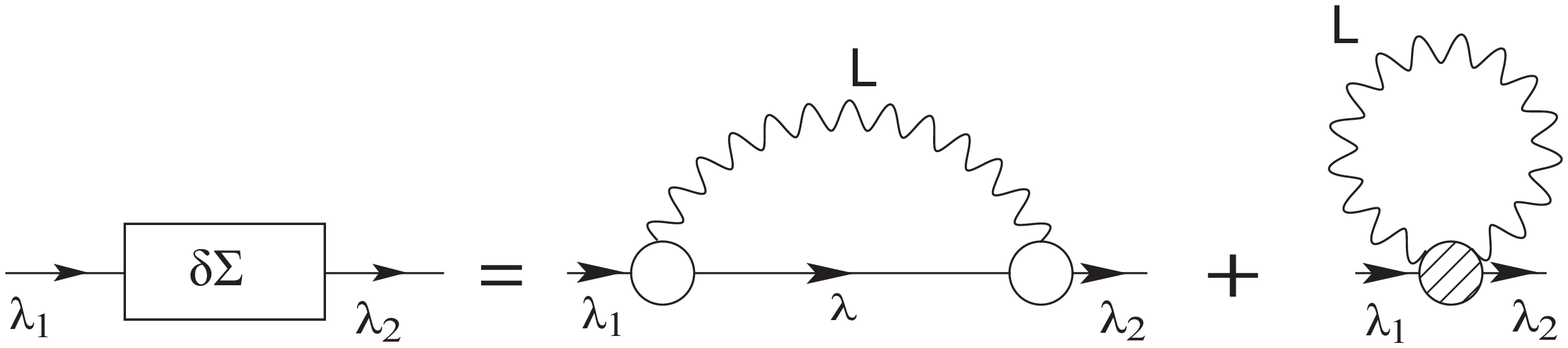}} \caption{Fig. 1. PC corrections to the mass
operator. The open circle is the vertex $g_L$ of the phonon creation. The gray blob denotes the phonon non-pole (``tadpole'') term.} \label{fig:SigPC}
\end{figure}

In magic nuclei, the perturbation theory in $\delta \Sigma^{\rm PC}_L$ is valid  \cite{levels} for
solving the Dyson equation with the mass operator $\Sigma(\eps){=}\Sigma_0{+}\delta \Sigma^{\rm
PC}(\eps)$. Another situation is often occurs in  semi-magic nuclei \cite{lev-semi}, due to a strong
mixture of some SP states with those possessing the structure of a SP state + $L$-phonon. For such
cases a method is developed in \cite{lev-semi} which is based on a direct solving the Dyson equation
with the mass operator $\Sigma(\eps)$. Each SP state $|\lambda\rangle$ splits to a set of
$|\lambda,i\rangle$ solutions with the SP strength distribution factors $S^i_{\lambda}$. In \cite
{lev-semi} a method is proposed how to express the average SP energy $\eps_{\lambda}$ and the average
$Z_{\lambda}$ factor in terms of  $\eps_{\lambda}^i$ and $S_{\lambda}^i$. It is similar to the one
used usually for finding the corresponding experimental values \cite{lev-exp}. The experimental data
in heavy non-magic nuclei considered in \cite{lev-semi} are practically absent, therefore no
comparison with experiment was made in that work.

Fortunately, there is a massive of data which has a direct relevance to the set of the solutions
$\eps_{\lambda}^i$ under discussion. We mean the odd-even mass differences, that is the ``chemical
potentials'' in the notation of the theory of finite Fermi systems (TFFS) \cite{AB}: \beq
\mu_+^n(Z,N)=-\left(B(Z,N+1)-B(Z,N)\right),\label{mun_pl}\eeq \beq
\mu_-^n(Z,N)=-\left(B(Z,N)-B(Z,N-1)\right).\label{mun_mi}\eeq \beq
\mu_+^p(Z,N)=-\left(B(Z+1,N)-B(Z,N)\right),\label{mup_pl}\eeq \beq
\mu_-^p(Z,N)=-\left(B(Z,N)-B(Z-1,N)\right),\label{mup_mi}\eeq where $B(Z,N)$ is the binding energy of
the corresponding nucleus. Evidently, they are equal to one nucleon separation energies $S_{n,p}$
\cite{BM1} taken with the opposite sign. For example, we have $\mu_-^n(Z,N){=}{-}S_n(Z,N)$  or
$\mu_+^n(Z,N){=}{-}S_n(Z,N+1)$.

Indeed, let us write down the Lehmann spectral expansion for the Green function $G(\eps;{\bf r_1},{\bf
r_2})$ in the $\lambda$-representation of the functions which diagonalize $G$ \cite{AB}: \beq
G_{\lambda}(\eps){=}\sum_{s} \frac {|(a^+_{\lambda})_{s0}|^2}{\eps{-}(E_s{-}E_0){+}i\gamma}
{+}\sum_{s} \frac {|(a_{\lambda})_{s0}|^2}{\eps{+}(E_s{-}E_0){-}i\gamma},\label{Lehmann}\eeq with
obvious notation. The isotopic index $\tau=(n,p)$ in (\ref{Lehmann}) is for brevity omitted. In both
the sums, the summation is carried out for the exact states $|s\rangle$ of nuclei with one added or
removed nucleon. Explicitly, if $|0\rangle$  is the ground state of the even-even ($Z,N$) nucleus, the
states $|s\rangle$ in the first sum correspond to the ($Z,N+1$) one for $\tau=n$ and ($Z+1,N$) for
$\tau=p$. Correspondingly, in the second sum they are ($Z,N-1$) for $\tau=n$ and (Z-1,$N$) for
$\tau=p$. If $|s\rangle$ is a ground state of the corresponding odd nucleus, the corresponding pole in
(\ref{Lehmann}) coincides with one the chemical potentials (\ref{mun_pl}) -- (\ref{mup_mi}). At the
mean field level, they can be attributed to the SP energies $\eps_{\lambda}$ with zero excitation
energy, whereas with account for the PC corrections they should coincide with the corresponding
energies $\eps_{\lambda}^i$.

The calculation scheme used in \cite{lev-semi} and in this work contains some approximations which are
typical for the TFFS \cite{AB} and which are valid in heavy nuclei with accuracy of $1/A$,
$A{=}N{+}Z$. At the mean field level, we relate the poles of the Green function $G(\eps)$ of the
even-even nucleus ($Z,N$) to the SP levels of its odd neighbors, ($Z\pm 1,N$) or ($Z,N\pm1$). Namely,
the effect of the core deformation by the odd particle or hole to the mass operator $\Sigma$ is
neglected. Note that, according to the TFFS scheme, this odd particle induced deformation and the
corresponding, say, quadrupole moment may be found explicitly by solving the equation for
the effective field [13], which is similar to that of the quasiparticle random phase approximation (QRPA), see e.g. \cite{BE2,Q-EPJA}. It should be stressed that the main odd particle effect
is the PC one, and we consider it explicitly. Another approximation we use at the PC stage and it concerns
the characteristics of the phonons we consider. Namely, we use the QRPA solution for the $L$-phonon in
the ($Z,N$) nucleus for finding the PC corrections to the SP characteristics of of these odd nuclei.
Thereby, we neglect the ``blocking effect'' of the odd particle (hole) in the QRPA equation for $g_L$.
Accuracy of this approximation can be estimated as $1/n_L$, where $n_L$ is the number of the
particle-hole states which contribute effectively to the vertex $g_L$. We deal with strongly
collective $2^+_1$ and $3^-_1$ states in the even lead isotopes, for which we may estimate this number as
$n_L{\simeq}20\div30$. A method to take into account the effect of the odd particle in the problem
under consideration is developed in \cite{Baldo-15}. It is mainly important for lighter nuclei. Note
also that there is a direct, in general more accurate, method to find the mass differences
(\ref{mun_pl}) -- (\ref{mup_mi}) in terms of the binding energies of each nuclei entering these
relations. However, the calculation of the PC corrections to binding energies is rather cumbersome
\cite{scTFFS} and up to now there is no systematic corresponding calculations.

Let us describe briefly the method \cite{lev-semi} to solve the PC corrected Dyson equation for the
quasiparticle Green function. We should solve the following equation: \beq \left(\eps-H_0 -\delta
\Sigma^{\rm PC}(\eps) \right) \phi =0, \label{sp-eq}\eeq where $H_0$ is the quasiparticle Hamiltonian
with the spectrum $\eps_{\lambda}^{(0)}$ and wave functions $\phi_{\lambda}^{(0)}$.

In the case when several $L$-phonons are taken into account, the total PC variation of the mass
operator in Eq. (\ref{sp-eq}) is  the sum over all phonons: \beq \delta \Sigma^{\rm PC} = \sum_L
\delta \Sigma^{\rm PC}_L . \label{sum-L}\eeq

We deal with the normal subsystem of the semi-magic nucleus under consideration, correspondingly,
$\Sigma(\eps)$ is the mass operator of a normal Fermi system. In this case, the explicit expression
for the pole term is well known and can be found in \cite{levels,lev-semi}. As to the non-local term,
we follow to the method developed  by Khodel \cite{Khod-76}, who first considered such diagrams in the
problem of PC corrections in nuclei, see also \cite{scTFFS}.

All low-lying phonons we deal with  are of surface nature,  the surface peak dominating in their
creation amplitude: \beq g_L(r)=\alpha_L \frac {dU} {dr} +\chi_L(r). \label{gLonr}\eeq The first term
in this expression  is surface peaked, whereas the in-volume addendum $\chi_L(r)$ is rather small. It
is illustrated in Fig. 2 for the $2^+_1$ and $3^-_1$ states in $^{198}$Pb. If one neglects this
in-volume term  $\chi_L$, very simple expression for the non-pole term can be obtained \cite{scTFFS}:
\beq \delta\Sigma^{\rm non-pole}_L = \frac {\alpha_L ^2} 2 \frac {2L+1} 3 \triangle U(r).
\label{tad-L}\eeq Just as in \cite{levels,lev-semi}, we will below neglect the in-volume term in
(\ref{gLonr}) and use Eq. (\ref{tad-L}) for the non-pole term of $\delta \Sigma_L^{\rm PC}$.

\begin{figure}
\centerline {\includegraphics [width=80mm]{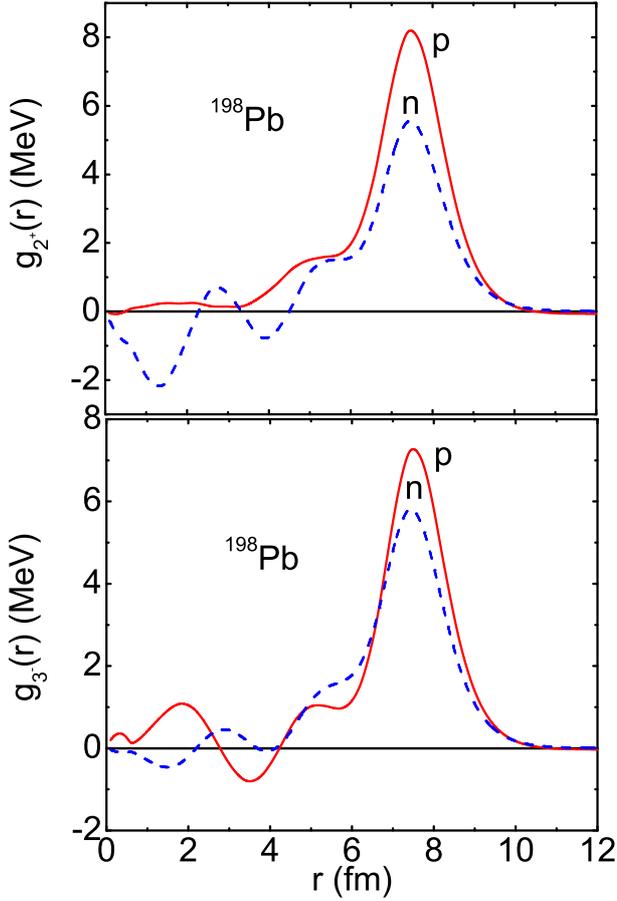}} \caption{Fig2. (Color online) Phonon creation
amplitudes $g_L(r)$ for two low-lying phonons in the $^{198}$Pb nucleus.} \label{fig:gL}
\end{figure}

\begin{figure}
\centerline {\includegraphics [width=80mm]{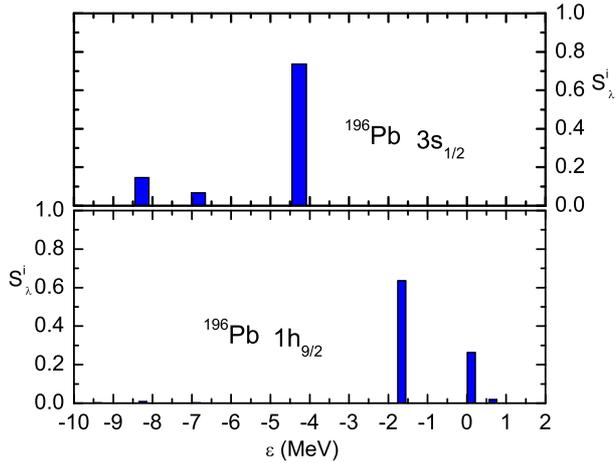}} \caption{Fig3. (Color online) The single-particle strengh distributions ($S$-factors) of two states nearby the Fermi level for $^{196}$Pb nucleus.}
\label{fig:Slam}
\end{figure}

In this work, we consider the chain of even lead isotopes, $^{180-214}$Pb, with account for two
low-lying phonons, $2^+_1$ and $3^-_1$. Their excitation energies $\omega_L$ and the coefficients
$\alpha_L$ in Eq. (\ref{gLonr}) are presented in Table 1. Comparison with existing experimental data
\cite{exp-omega} is given. We present only 3 decimal signs only of the latter to avoid a
cumbersomeness of the table. On the whole, the $\omega_L$ values agree with the data sufficiently
well. In more detail, for the interval of $^{194-200}$Pb, the theoretical excitation energies of the
$2^+$-states are visibly less than the experimental ones. This is a signal of the fact that our
calculations overestimate the collectivity of these states and, correspondingly, the PC effect in
these nuclei. The opposite situation where is for the lightest Pb isotopes, $A{<}190$, where we,
evidently, underestimate the PC effect. The $\alpha_L$ value defines the amplitude, directly in fm, of
the surface $L$-vibration in the nucleus under consideration. We see that in the most cases both the
phonons we consider are strongly collective, with $\alpha_L{\simeq}0.3$ fm. At small values of
$\omega_L$, the vibration amplitude behaves as $\alpha_L {\sim}1/\omega_L$ \cite{scTFFS,BM2}.   Both
the PC corrections to the SP energy, pole and non-pole, are proportional to $\alpha_L^2$.   The ghost
state $1^-$ is also taken into account, although the corresponding correction for nuclei under
consideration is very small, see \cite{lev-semi}, because it depends on the mass number as $1/A$,
$A{=}N{+}Z$.

\begin{table}[h!]
\caption{Table 1. Excitation energies  $\omega_L$ (MeV) and the coefficients $\alpha_L$ (fm) in Eq.
(\ref{gLonr}) of the $2^+_1$ and $3^-_1$ phonons in even Pb isotopes.}
\begin{tabular}{c| c| c| c| c |c |c }

\hline\noalign{\smallskip} A  & $\omega_2$ & $\omega_2^{\rm exp}$  & $\alpha_2$
& $\omega_3$ & $\omega_3^{\rm exp}$    & $\alpha_3$    \\
\noalign{\smallskip}\hline\noalign{\smallskip}

180  &   1.415 & 1.168(1) &0.31 &  2.008 & -- &0.35 \\
182  &   1.284 &0.888  &0.31 &  1.836 & -- &0.35 \\
184  &   1.231 & 0.702 &0.32 &  1.839 & -- &0.36 \\
186  &   1.133 & 0.662 &0.33 &  1.881 & -- &0.34 \\
188  &   1.028 & 0.724 &0.34 &  1.968 & -- &0.34 \\
190  &   0.930 & 0.774 &0.36 &  2.052 & -- &0.33 \\
192  &   0.849 &0.854  &0.35 &  2.160 & -- &0.32 \\
194  &   0.792 & 0.965 &0.35 &  2.272 & -- &0.32 \\
196  &   0.764 & 1.049 &0.35 &  2.390 & 2.471(?) &0.31 \\
198  &   0.762 & 1.064 &0.35 &  2.506 & -- &0.31 \\
200  &   0.789 & 1.027 &0.30 &  2.620 & --     &0.31 \\
202  &   0.823 & 0.961 &0.31 &  2.704 & 2.517 &0.31 \\
204  &   0.882 & 0.899 &0.22 &  2.785 & 2.621 &0.31 \\
206  &   0.945 & 0.803 &0.16 &  2.839 & 2.648 &0.32 \\
208  &   4.747 & 4.086 &0.33 &  2.684 &2.615  &0.09 \\
210  &   1.346 & 0.800 &0.07 & 2.183  & 1.870(10) &0.19 \\
     &         &  &     & 2.587  &2.828(10)  &0.17  \\
212  &   1.444 & 0.805 &0.17 & 1.788  & 1.820(10) &0.36  \\
214  &   1.125 & 0.835(1) &0.19 & 1.469  & -- &0.37  \\

\noalign{\smallskip}\hline

\end{tabular}\label{tab1}
\end{table}

\newpage

\begin{table}[]
\caption{Table 2. Examples of solutions of Eq. (\ref{sp-eq1}) for protons in the $^{198}$Pb nucleus.}
\begin{tabular}{ c| c| c | l }

\hline\noalign{\smallskip}

$\lambda$   & $i$ & $\eps_{\lambda}^i$, MeV  &\hspace*{7mm} $S_{\lambda}^i$   \\

\noalign{\smallskip}\hline\noalign{\smallskip}
3$s_{1/2}$  &  1  &    -8.701 &  0.144 \\
            &  2  &    -7.270 &  0.667 $\times 10^{-1}$\\
            &  3  &    -4.716 &  0.741 \\
            &  4  &     2.078 &  0.194$\times 10^{-1}$ \\
\noalign{\smallskip}\hline\noalign{\smallskip}
            &     &   &$\sum S_{\lambda}^i{=}0.970$  \\
\noalign{\smallskip}\hline\noalign{\smallskip}
1$h_{9/2}$ &  1  &   -11.974  &  0.250 $\times 10^{-2}$ \\
           &  2  &    -9.952  &  0.895 $\times 10^{-3}$ \\
           &  3  &    -8.795  &  0.749 $\times 10^{-2}$ \\
           &  4  &    -7.357  &  0.127 $\times 10^{-2}$ \\
           &  5  &    -2.212  &  0.636  \\
           &  6  &    -0.466  &  0.272  \\
           &  7  &     0.199  &  0.139 $\times 10^{-1}$ \\
           &  8  &     2.750  &  0.481 $\times 10^{-2}$ \\
\noalign{\smallskip}\hline\noalign{\smallskip}
           &     &            &$\sum S_{\lambda}^i{=}0.939$ \\

 \noalign{\smallskip}\hline

\end{tabular}\label{tab2}
\end{table}

\begin{table*}
\caption{Table 3. Proton odd-even mass differences $\mu_{\pm}$ MeV for the even Pb isotopes. The mean
field predictions for the Fayans EDF DF3-a and those with the PC corrections are given.}
\begin{tabular}{c|c|c|c|c|c}
\hline

nucl.& $\lambda$ & DF3-a &  DF3-a + $2^+$  & DF3-a + $(2^+{+}3^-)$  & exp \cite{mass}\\
\hline

$^{180}$Pb& $\mu_+$, 1h$_{9/2}$ &  3.513   &  3.185 &   3.321 &    ---     \\
          & $\mu_-$, 3s$_{1/2}$ & -1.119   & -0.571 &  -0.793 &   -0.938(0.054) \\

$^{182}$Pb& $\mu_+$, 1h$_{9/2}$ &  2.942   &  2.564 &   2.695 &    --- \\
          & $\mu_-$, 3s$_{1/2}$ &  -1.610  & -1.023 &  -1.268 &   -1.316(0.021) \\

$^{184}$Pb& $\mu_+$, 1h$_{9/2}$ &  2.360   &  1.906 &   2.093 &    1.527(0.094) \\
          & $\mu_-$, 3s$_{1/2}$ & -2.104   & -1.450 &  -1.727 &   -1.753(0.022) \\

$^{186}$Pb& $\mu_+$, 1h$_{9/2}$ & 1.767    &  1.293 &   1.441 &    1.010(0.021) \\
          & $\mu_-$, 3s$_{1/2}$ & -2.592   & -1.906 &  -2.152 &   -2.213(0.032) \\

$^{188}$Pb& $\mu_+$, 1h$_{9/2}$ & 1.172    &  0.683 &   0.806 &    0.461(0.031) \\
          & $\mu_-$, 3s$_{1/2}$ &-3.072    & -2.356 &  -2.561 &   -2.661(0.019) \\

$^{190}$Pb& $\mu_+$, 1h$_{9/2}$ & 0.577    &  0.027 &   0.141 &   -0.112(0.020) \\
          & $\mu_-$, 3s$_{1/2}$ & -3.543   & -2.750 &  -2.945 &   -3.103(0.023) \\

$^{192}$Pb& $\mu_+$, 1h$_{9/2}$ & -0.017   & -0.528 &  -0.420 &   -0.596(0.022) \\
          & $\mu_-$, 3s$_{1/2}$ & -4.005   & -3.265 &  -3.440 &   -3.572(0.020) \\

$^{194}$Pb& $\mu_+$, 1h$_{9/2}$ & -0.608   & -1.167 &  -1.058 &   -1.107(0.023) \\
          & $\mu_-$, 3s$_{1/2}$ & -4.461   & -3.673 &  -3.838 &   -4.019(0.024) \\

$^{196}$Pb& $\mu_+$, 1h$_{9/2}$ & -1.193   & -1.760 &  -1.658 &   -1.615(0.023) \\
          & $\mu_-$, 3s$_{1/2}$ & -4.911   & -4.111 &  -4.268 &   -4.494(0.025) \\

$^{198}$Pb& $\mu_+$, 1h$_{9/2}$ & -1.769   & -2.316 &  -2.212 &   -2.036(0.025) \\
          & $\mu_-$, 3s$_{1/2}$ & -5.358   & -4.569 &  -4.716 &   -4.999(0.031) \\

$^{200}$Pb& $\mu_+$, 1h$_{9/2}$ & -2.327   & -2.757 &  -2.648 &   -2.453(0.026) \\
          & $\mu_-$, 3s$_{1/2}$ & -5.806   & -5.177 &  -5.317 &   -5.480(0.039) \\

$^{202}$Pb& $\mu_+$, 1h$_{9/2}$ & -5.806   & -5.177 &  -5.317 &   -5.480(0.039) \\
          & $\mu_-$, 3s$_{1/2}$ & -6.258   & -5.612 &  -5.753 &   -6.050(0.018) \\

$^{204}$Pb& $\mu_+$, 1h$_{9/2}$ & -3.356   & -3.567 &  -3.447 &   -3.244(0.006) \\
          & $\mu_-$, 3s$_{1/2}$ & -6.717   & -6.357 &  -6.493 &   -6.637(0.003) \\

$^{206}$Pb& $\mu_+$, 1h$_{9/2}$ & -3.818   & -3.911 &  -3.771 &   -3.558(0.004) \\
          & $\mu_-$, 3s$_{1/2}$ & -7.179   & -6.976 &  -7.105 &   -7.254(0.003) \\

$^{208}$Pb& $\mu_+$, 1h$_{9/2}$ & -4.232   & -4.064 &  -3.959 &   -3.799(0.003) \\
          & $\mu_-$, 3s$_{1/2}$ & -7.611   & -7.778 &  -7.633 &   -8.004(0.007) \\

$^{210}$Pb& $\mu_+$, 1h$_{9/2}$ & -4.670   & -4.653 &  -4.566 &   -4.419(0.007) \\
          & $\mu_-$, 3s$_{1/2}$ & -8.030   & -7.971 &  -8.055 &   -8.379(0.010) \\

$^{212}$Pb& $\mu_+$, 1h$_{9/2}$ & -5.111   & -5.152 &  -4.980 &   -4.972(0.007) \\
          & $\mu_-$, 3s$_{1/2}$ & -8.446   & -8.276 &  -8.481 &   -8.758(0.044) \\

$^{214}$Pb& $\mu_+$, 1h$_{9/2}$ & -5.555   & -5.686 &  -5.523 &   -5.460(0.017) \\
          & $\mu_-$, 3s$_{1/2}$ & -8.857   & -8.620 &  -8.865 &   -9.254(0.029) \\
\hline

 &$\langle \delta \mu \rangle_{\rm rms}$ & 0.385  & 0.321 &  0.218& \\

\hline

\end{tabular}\label{tab3}
\end{table*}

As the non-regular PC corrections to the SP energies we examine are important only for the states
nearby the Fermi surface, we limit ourselves with a model space $S_0$ including two shells close to
it, i.e., one hole and one particle shells, and besides we retain only the negative energy states.
Note that for finding the pole term $\delta \Sigma^{\rm pole}$ we use essentially wider SP space with
energies $\eps^{(0)}_{\lambda}{<}40$ MeV. To illustrate the method, we take  for example the nucleus
$^{198}$Pb . The space $S_0$ involves 5 hole states ($1g_{7/2}$, $2d_{5/2}$, $1h_{11/2}$, $2d_{3/2}$,
$3s_{1/2}$) and three particle ones ($1g_{9/2}$, $2f_{7/2}$, $1i_{13/2}$). We see that there is here
only one state for each $(l,j)$ value. Therefore, we need only diagonal elements of $\delta
\Sigma^{\rm PC}$ in Eq. (\ref{sp-eq}). In the result, it reduces as follows:
 \beq \eps-\eps_{\lambda}^{(0)}-\delta \Sigma^{\rm PC}_{\lambda \lambda}(\eps) =0.
\label{sp-eq1}\eeq

Details of finding the solutions $\eps_{\lambda}^i$ of Eq. (\ref{sp-eq1})  can be found in
\cite{lev-semi}. In this notation, $\lambda$ is just the index for the initial SP state from which the
state $|\lambda,i\rangle$ originated. The corresponding SP strength distribution factors ($S$-factors)
are: \beq S_{\lambda}^i =\left(1- \left(\frac {\partial} {\partial \eps}
 \delta \Sigma^{\rm PC}(\eps) \right)_{\eps=\eps_{\lambda}^{i}}\right)^{-1}. \label{S-fac}\eeq
They should obey the normalization rule: \beq \sum_i S_{\lambda}^i =1.   \label{norm}\eeq Accuracy of
fulfillment of this relation is a measure of the completeness of the model space $S_0$ we use to solve
the problem under consideration. Two examples of the sets of solutions for four $|\lambda,i\rangle$
states in $^{198}$Pb are presented in Table 2. They originate from the first hole and the first
particle states in the model space $S_0$. In this case, our prescription for the odd-even mass differences, in accordance with the Lehmann
expansion (\ref{Lehmann}), is as follows: \beq \mu_+(^{198}{\rm Pb})=-2.212\,\, {\rm MeV}, \eeq \beq
\mu_-(^{198}{\rm Pb})=-4.716\,\, {\rm MeV}. \eeq As an illustration, we displayed in Fig. 3 the SP strength distributions ($S$-factors) of the similar two states nearby the Fermi level in the neighboring nucleus
$^{196}$Pb.

The similar values for all the chain under consideration are given in Table 3. In the last line, the
average deviation is given of the theoretical predictions from existing experimental data: \beq
\langle \delta \mu \rangle_{\rm rms} = \sqrt{ \sum (\mu_{\pm}^{\rm th}-\mu_{\pm}^{\rm exp})^2/N_{\rm
exp}}, \label{rms}\eeq where $N_{\rm exp}{=}34$. For comparison, we calculated the corresponding value
for the ``champion'' Skyrme EDF HFB-17 \cite{HFB-17} using the table \cite{HFB-site} of the nuclear
binding energies. It is equal to $\langle \delta \mu \rangle_{\rm rms}({\rm HFB-17}){=}0.333$ MeV. We
see that accuracy of the Fayans EDF DF3-a without PC in predicting the odd-even mass differences is
only a bit worse. It agrees with the original Fayans's idea \cite{Fay1,Fay} develop an EDF without PC
corrections. However, account for the PC corrections due to two low-laying collective phonons makes
agreement with the data significantly better.

To resume, a method, developed recently \cite{lev-semi} to find the PC corrections to SP energies of
semi-magic nuclei based on the direct solution of the Dyson equation with the PC corrected mass
operator, is used for finding the odd-even mass difference between the even Pb isotopes and their
odd-proton neighbors. The Fayans EDF DF3-a is used for generating the mean field basis. On the
mean-field level, the average accuracy of the predictions for the mass differences $\langle \delta \mu
\rangle_{\rm rms}({\rm DF3{-}a}){=}0.389$ MeV is only a bit worse than that (0.333 MeV) for the Skyrme
EDF HFB-17 fitted to nuclear masses with highest accuracy among the self-consistent calculations.
Account for the PC corrections due to the low-laying phonons $2^+_1$ and $3^-_1$ makes the agreement
significantly better, $\langle \delta \mu \rangle_{\rm rms}({\rm DF3{-}a})^{\rm PC}{=}0.218$ MeV.

\vskip 0.3 cm We acknowledge for support the Russian Science Foundation, Grants Nos. 16-12-10155 and
16-12-10161. The work was also partly supported  by the RFBR Grant 16-02-00228-a. This work was
carried out using computing resources of federal center for collective usage at NRC ``Kurchatov
Institute'', http://ckp.nrcki.ru. EES thanks the Academic Excellence Project of the NRNU MEPhI under
contract by the Ministry of Education and Science of the Russian Federation No. 02. A03.21.0005.

{}

\end{document}